\documentclass[aps,prl,twocolumn,showpacs,superscriptaddress,preprintnumbers,amsmath,amssymb]{revtex4}

\usepackage{amsmath}
\usepackage{graphicx}
\usepackage{bm}
\usepackage{amsfonts}
\usepackage{amssymb}
\usepackage{multirow}
\usepackage{subfigure}
\usepackage[usenames,dvipsnames]{color}

\begin{document}

\title{Experimental demonstration of an inertial collimation mechanism in nested outflows }

\author{R. Yurchak}\affiliation{LULI, Ecole Polytechnique, CNRS, CEA, UPMC, Route de Saclay, 91128 Palaiseau, France}
\author{A. Ravasio}\thanks{alessandra.ravasio@polytechnique.fr}
\affiliation{LULI, Ecole Polytechnique, CNRS, CEA, UPMC, Route de Saclay, 91128 Palaiseau, France}
\author{A. Pelka}\affiliation{LULI, Ecole Polytechnique, CNRS, CEA, UPMC, Route de Saclay, 91128 Palaiseau, France}
\author{S. Pikuz Jr.}\affiliation{Joint Institute for High Temperatures RAS, 13-2 Izhorskaya st., Moscow, 125412 Russia.}
\author{E. Falize}  \affiliation{CEA-DAM-DIF, F-91297 Arpajon, France}
\author{T. Vinci} \affiliation{LULI, Ecole Polytechnique, CNRS, CEA, UPMC, Route de Saclay, 91128 Palaiseau, France}
\author{M. Koenig}\affiliation{LULI, Ecole Polytechnique, CNRS, CEA, UPMC, Route de Saclay, 91128 Palaiseau, France}
\author{B. Loupias}\affiliation{CEA-DAM-DIF, F-91297 Arpajon, France}
\author{A. Benuzzi-Mounaix}\affiliation{LULI, Ecole Polytechnique, CNRS, CEA, UPMC, Route de Saclay, 91128 Palaiseau, France}
\author{M. Fatenejad}\affiliation{Flash Center for Computational Science, University of Chicago, IL 60637}
\author{P. Tzeferacos}\affiliation{Flash Center for Computational Science, University of Chicago, IL 60637}
\author{D. Q. Lamb}\affiliation{Flash Center for Computational Science, University of Chicago, IL 60637}
\author{E.G. Blackman}\affiliation{Department of Physics and Astronomy, University of Rochester, Rochester, NY 14627, USA}

\date{\today}

\begin{abstract} 

Interaction between a central outflow and  a surrounding  wind is   common in astrophysical sources  powered by accretion.
Understanding  how the interaction might help to collimate the inner  central outflow  is of  interest for assessing astrophysical jet formation paradigms. In this context, we studied the interaction between two nested supersonic plasma flows generated by focusing a long pulse high-energy laser beam onto a solid target. A nested geometry was created by shaping the energy distribution at the focal spot with a dedicated phase plate. Optical and X-ray diagnostics were used to study the interacting flows. Experimental results and  numerical hydrodynamic simulations  indeed show the formation of  strongly collimated jets. 
Our work  experimentally confirms  the ``shock-focused inertial confinement"  mechanism proposed in previous theoretical astrophysics investigations.
 \end{abstract}

\maketitle
\newpage



{\it Introduction}-Supersonic jets are common in astrophysics, emanate from such sources as newly forming young stellar objects  (YSOs) \cite{reipurth01}, active galactic nuclei (AGN) \cite{Ferrari1998,Marscher2006}, 
planetary and pre-planetary nebulae  (PPN and PPN) \cite{Balick} and  micro-quasars \cite{mirabel}.
Their sustained collimation over large distances is not yet completely understood. Both 
magnetohydrodynamic (MHD) and hydrodynamic (HD) processes may be important.
 Often the jets  propagate  within a surrounding wind or envelope, as observed in YSOs   \cite{arce02},
  AGN \cite{Tombesi2012}, and  in  PN, where fast collimated winds sweep into a slower denser wind ejected most strongly during the PPN phase \cite{bujarrabal,rizzo2013}.    
   For YSOs and AGN  a direct connection between disks and jets has been established and there is emerging consensus for such in the PPN/PN context as well \cite{soker98,witt09,blackmanlucchini}.  
The question of how  different   time-dependent ambient thermal and ram pressures affect jet collimation arises quite generally \cite{icke93,Frank1994,Frank1996,BlackmanApJ2004, DennisApJ2009}.\\
\indent
The role of the ambient medium can be important even if the inner outflows are
 magnetically driven \cite{Fendt,lb03}. Recent 3D MHD simulations of  laser driven plasma experiments have looked at the possible magnetic field collimation of wide-angle winds into HD jets \cite{Ciardi} and interpreted this as analogous to hydrodynamic collimation of an inner flow by a  torus.  
	Astrophysical jet launch regions are generally not observationally resolved, being obscured by high opacities. It is there- fore valuable to distill the distinct physics of MHD and HD effects via alternative methods.\\
\indent
Combined with numerical simulations and theory,  experiments bring new contributions to the subject. 
	Some    jet  propagation and collimation mechanisms within steady ambient backgrounds have been studied experimentally  \cite{Loupias2007, Farley1999, Lebedev2002}. Crosswinds were also used to study jet deflection and C-shape structures \cite{Lebedev2004}. 
Here we present results from  a new  experimental approach aimed at   investigating the time dependent HD collimation of an inner isotropically supersonic expanding plasma by a surrounding time-evolving  supersonic ambient flow. 

\begin{figure}[t]
\begin{center}
\includegraphics[scale=0.5]{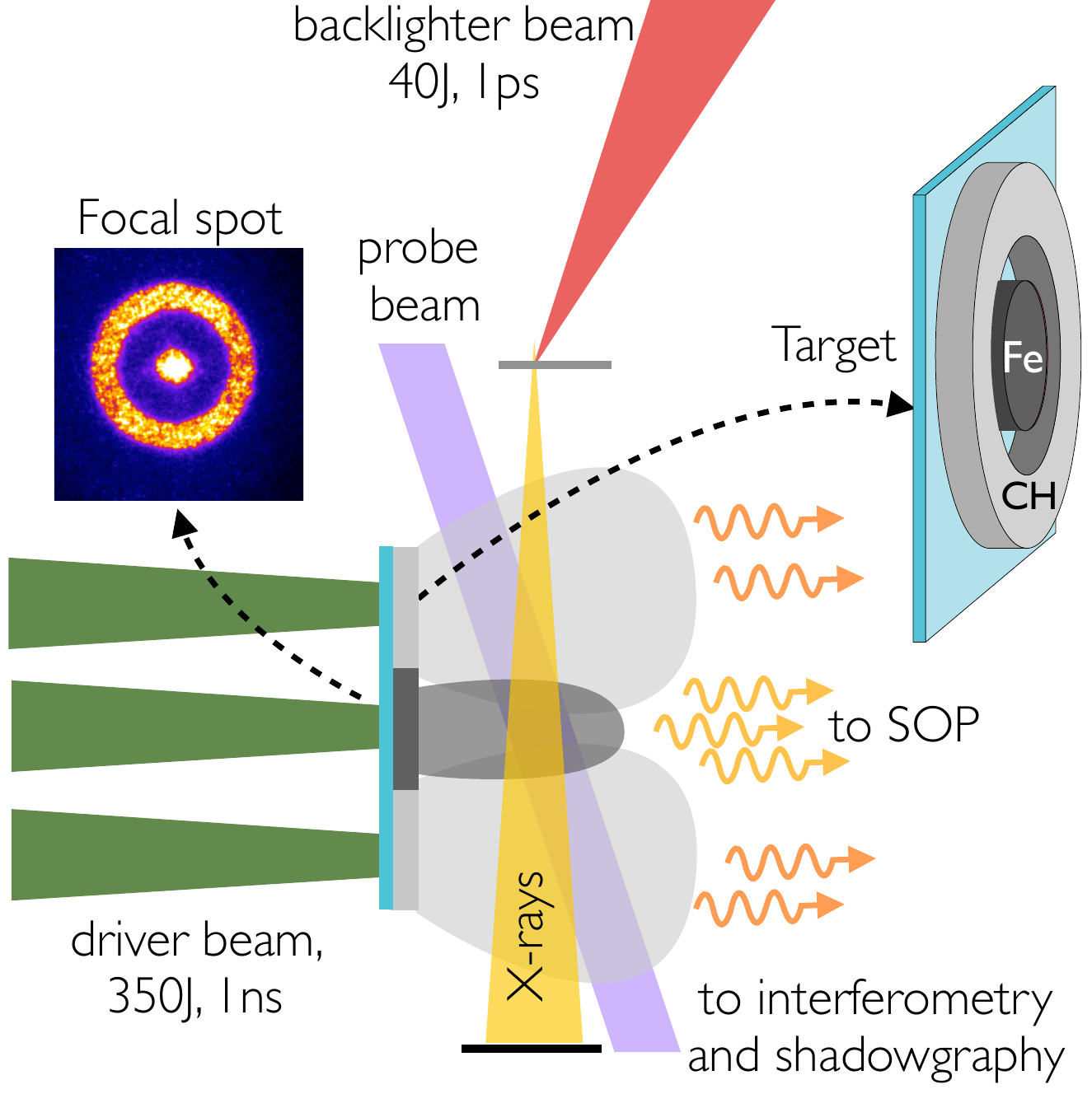} 
\caption{Schematic of the experimental setup, focal spot (measured) and target geometry. The laser energy distribution is modeled with a hybrid phase plate (HPP) resulting in a central disc and an outer ring.  The nested target fits this geometry, consisting in a central iron (Fe 15$\mu$m) disc and and outer plastic (CH 15$\mu$m) ring on a common CH-Al pusher. 
} 
\label{fig: setup}
\end{center} 
\end{figure}

\begin{figure}[t]
\begin{center}
\includegraphics[scale=0.35]{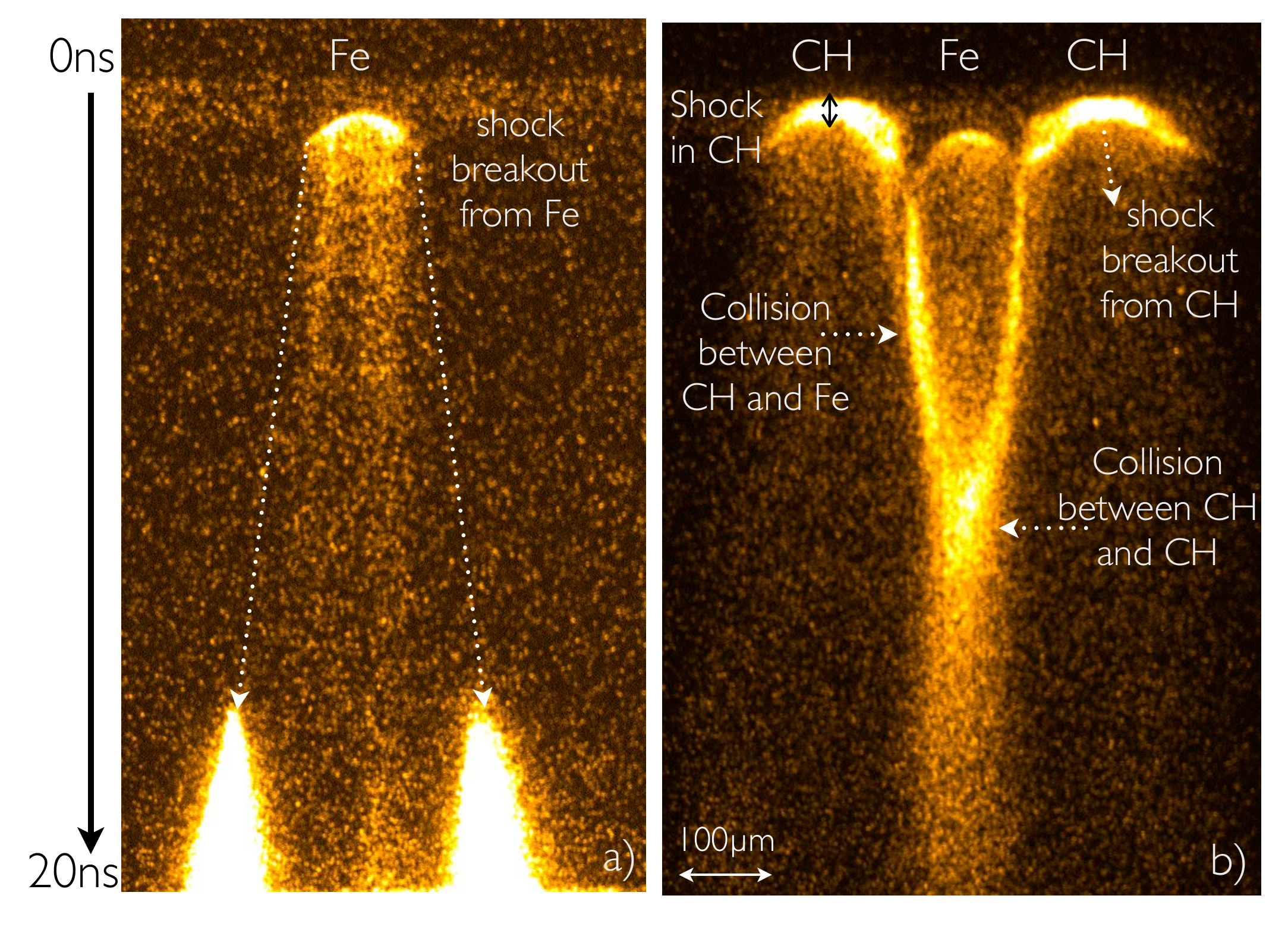} 
\caption{Typical time resolved rear side self-emission data for a) the iron disc only and b) the complete nested target.} 
\label{fig: SOP1D} 
\end{center} 
\end{figure}

{\it Experimental Setup-}The experiment was performed on the LULI2000 laser facility at the LULI Laboratory, in France. The set-up is shown schematically in Fig. \ref{fig: setup}. A long pulse ($\tau_L\sim$ 1.5 ns) high-energy  ($E_L\sim$ 400 J at $\lambda_L=$ 527 nm) laser beam was used to produce supersonic plasma flows via  interaction with solid targets.
To create the nested configuration, we have designed a phase plate able to generate a laser energy distribution with a (100 $\mu$m) central circular spot and a thinner (75 $\mu$m) outer ring. 
Targets manufactured to match this pattern were made of a central iron disk (15 $\mu$m thick Fe) and a peripheral plastic ring (CH, also 15 $\mu$m thick), sitting on a CHAl pusher.
Upon laser impact, a shock wave is launched in the pusher and transmitted to the Fe and CH layers. Once this shock reaches the rear side of the target, supersonic plasmas are formed from the outer plastic ring and from the central iron disk.\\
\indent
 To probe the interacting flows,  
 we used rear-side and transverse optical diagnostics, in addition to transverse X-ray radiography.
Optical probes were applied to the low density CH plasma and X-rays were used to characterize the inner iron flow which is opaque to optical radiation.
Transverse optical diagnostics included time-resolved self-emission, shadowography and interferometry, while time-resolved 1D self-emission and 2D self-emission snapshots were implemented at the rear side of the target. This ensemble allowed us to measure the plastic flow velocity, morphology and electron density.\\
\indent
{\it Results-}
  Typical data showing
  time resolved 1-D self-emission are
  shown in Fig. \ref{fig: SOP1D},  
 In Fig.  \ref{fig: SOP1D}a, the ring part of the laser spot was blocked and the data show a typical plasma release, with the iron expanding and cooling into vacuum.  
Fig. \ref{fig: SOP1D}b shows the case of a complete target, wtih the plastic ring added. Since the plastic is transparent to  visible light, we can follow the shock front in the plastic layer, and measure the shock velocity \textit{D}. Typical \textit{D} values are $\sim$30 km/s. When the shock breaks out, the CH unloads into vacuum at $\sim$70 km/s, as measured from the transverse time-resolved self-emission. The shock wave propagates  more slowly in iron than  plastic because of different impedances. Therefore the Fe flow unloads into vacuum after the CH 
and  then swiftly collides with the radially expanding CH.
Bright emission is observed from this collision and it is associated to a shock generation, as we will discuss below. 
The time evolution of the collision emission shows that the iron 
is confined by the surrounding CH flow. Later, the CH flows themselves collapse on  axis, generating highly collimated emission.

\begin{figure}
\begin{center}
\includegraphics[width=1\textwidth]{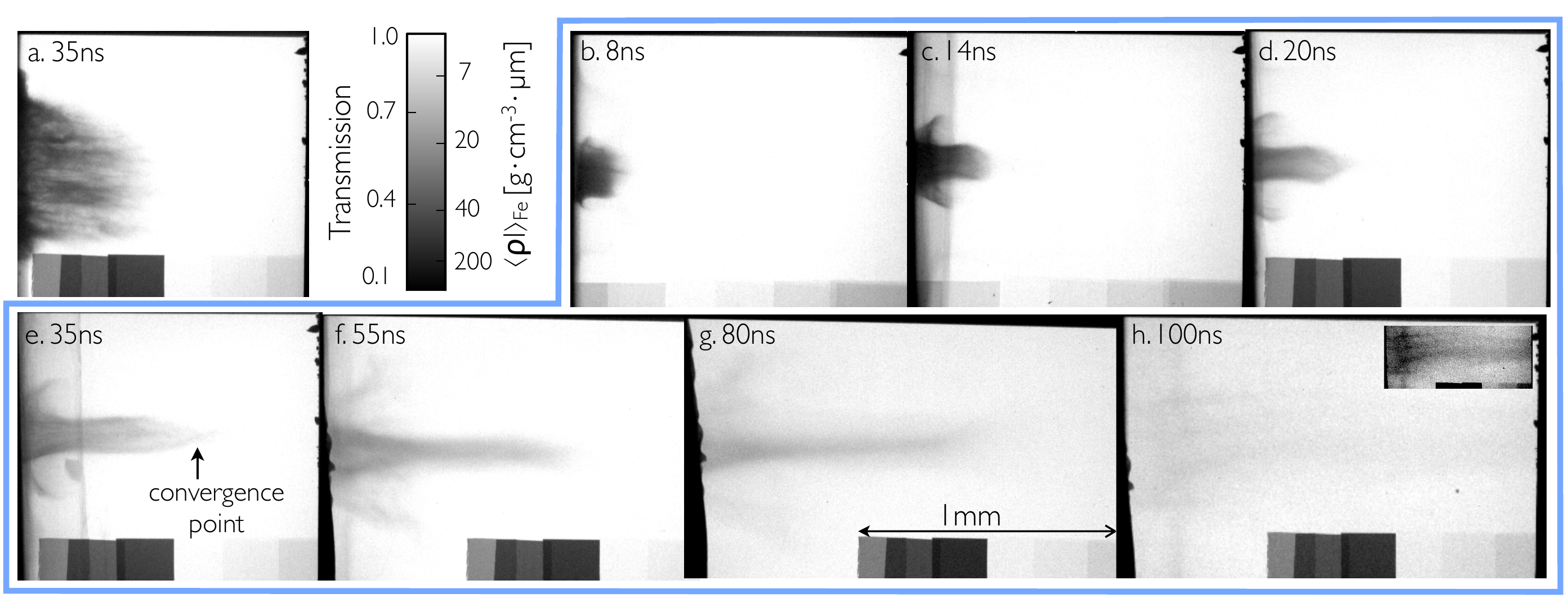}
\caption[Radiography time series]{X-ray radiography of the iron flows from a) the central disk alone, where the outer ring was blocked on the laser side and b)-h) from the complete nested target 
In this case, we see the formation of a collimated flow being stable up to very long delays (80 ns) and reaching mm sizes. In the insertion at 100 ns  we have enhanced the contrast.  
\label{radiography:time_series}}
\end{center}
\end{figure}

  X-ray radiography confirms the iron collimation by the plastic flows (Fig. \ref{radiography:time_series}). The X-ray source was generated by driving a copper back-lighter with the short pulse beam ($\sim$1 ps) of the pico2000 laser system. The incident X-ray spectrum 
accounted for  intense K-$\alpha$ emission line at $\sim$8 keV, superimposed on a  weaker bremsstrahlung continuum. 
At these X-ray energies, the outer plastic plasma is nearly transparent, while the iron flow is highly absorbing. The central (iron) jet morphology can therefore be identified without being disturbed by the surrounding flow. 
Again we have compared shots taken with and without the surrounding plastic.
Typical results are presented in figure \ref{radiography:time_series}a and \ref{radiography:time_series}e for a probing delay of 35 ns.
As in  Fig. \ref{fig: SOP1D}a, the case of iron alone (Fig. \ref{radiography:time_series}a) exhibits quasi-spherical adiabatic expansion. As soon as the plastic flow is added, the expansion is strongly reduced and the flow is collimated. Fig. \ref{radiography:time_series}e thus  confirms what the optical data suggested more indirectly.  
By varying the delay of the back-lighter to the main beam from 8\,ns to 100\,ns, we  monitored the time evolution of the iron flow. The results are presented in Fig.\ref{radiography:time_series}b-h,  confirming the emergence of a  thin jet from an initially uncollimated plasma. 

Overall, the data of Fig. \ref{radiography:time_series} reveal different phases of jet evolution. First, the iron expands. A high density layer is formed at the interface between the iron and plastic plasma, detected as a reduced  transmission at the Fe boundary  in the radiographs of figure \ref{radiography:time_series}c.  The plasma flow is subsequently focused on the axis, with a convergence point clearly observable at 35 ns (Fig. \ref{radiography:time_series}e). At longer times, a  narrow collimated feature is observed remaining stable for  80 ns (figure \ref{radiography:time_series}g).\\
  The jet longitudinal extension (i.e. along the propagation axis) linearly increases from a few 100 $\mu$m at earlier times (Fig. \ref{radiography:time_series}b-e) to mm scales at later times (Fig. \ref{radiography:time_series}f-g). Its radius  shrinks in time  and can be fit by the expression $r(\mu$m)$\sim$56$\cdot$e$^{-t (ns)/13}$+53. 
The iron confinement sustains high surface densities. From transmission data we measure 100 g$\cdot$ cm$^{-3}\cdot \mu$m at earlier times (8 ns) and 12g$\cdot$ cm$^{-3}\cdot \mu$m at 80 ns. The corresponding densities at the mid length of the jet are obtained by  Abel inversion resulting in 0.6 g/cc and 0.1 g/cc respectively. At these densities, typical aspect ratios (AR, length-to-width) up to $\sim$5 are obtained. 

\begin{figure}
\begin{center}
\includegraphics[width=1\textwidth]{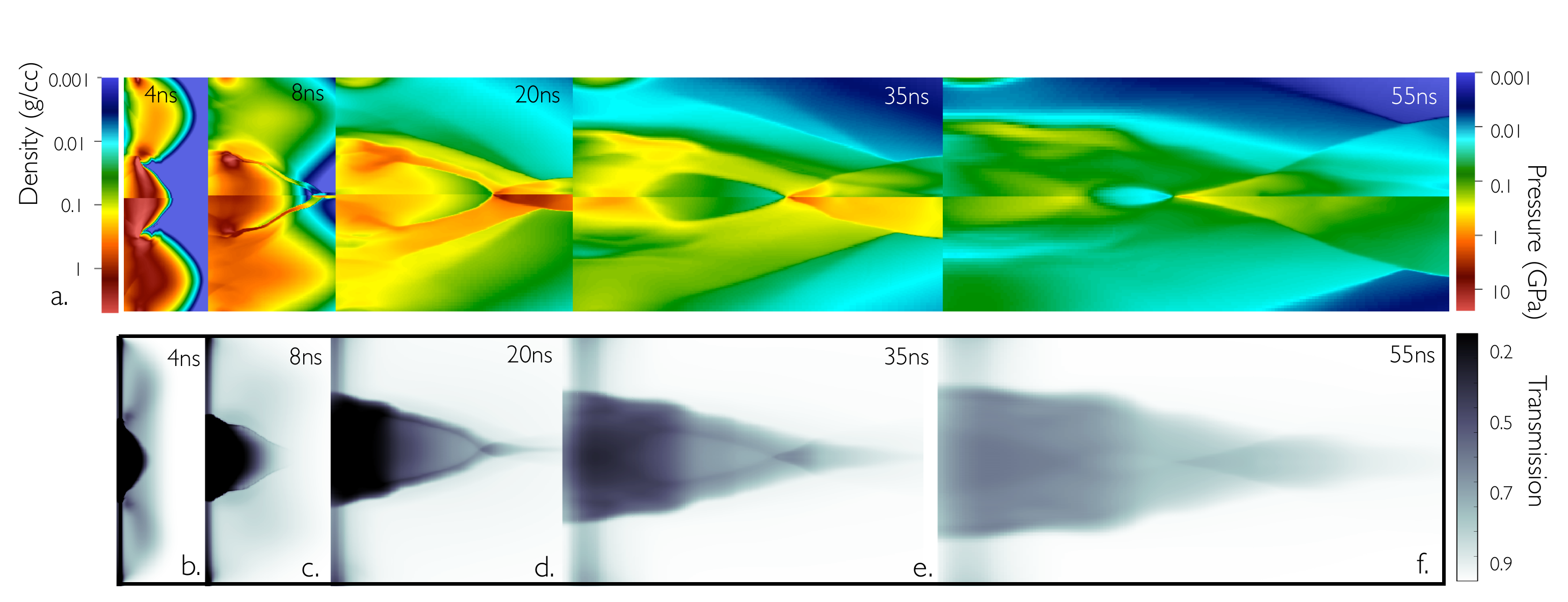}
\caption{Upper panel: simulated density (upper) and pressure (lower) maps at different time delays. Lower panel: corresponding synthetic radiographs obtained using the simulated density map and the experimental X-ray spectrum. \label{Simulations}}
\end{center}
\end{figure}
\begin{figure}[t]
\begin{center}
\includegraphics[width=0.5\textwidth]{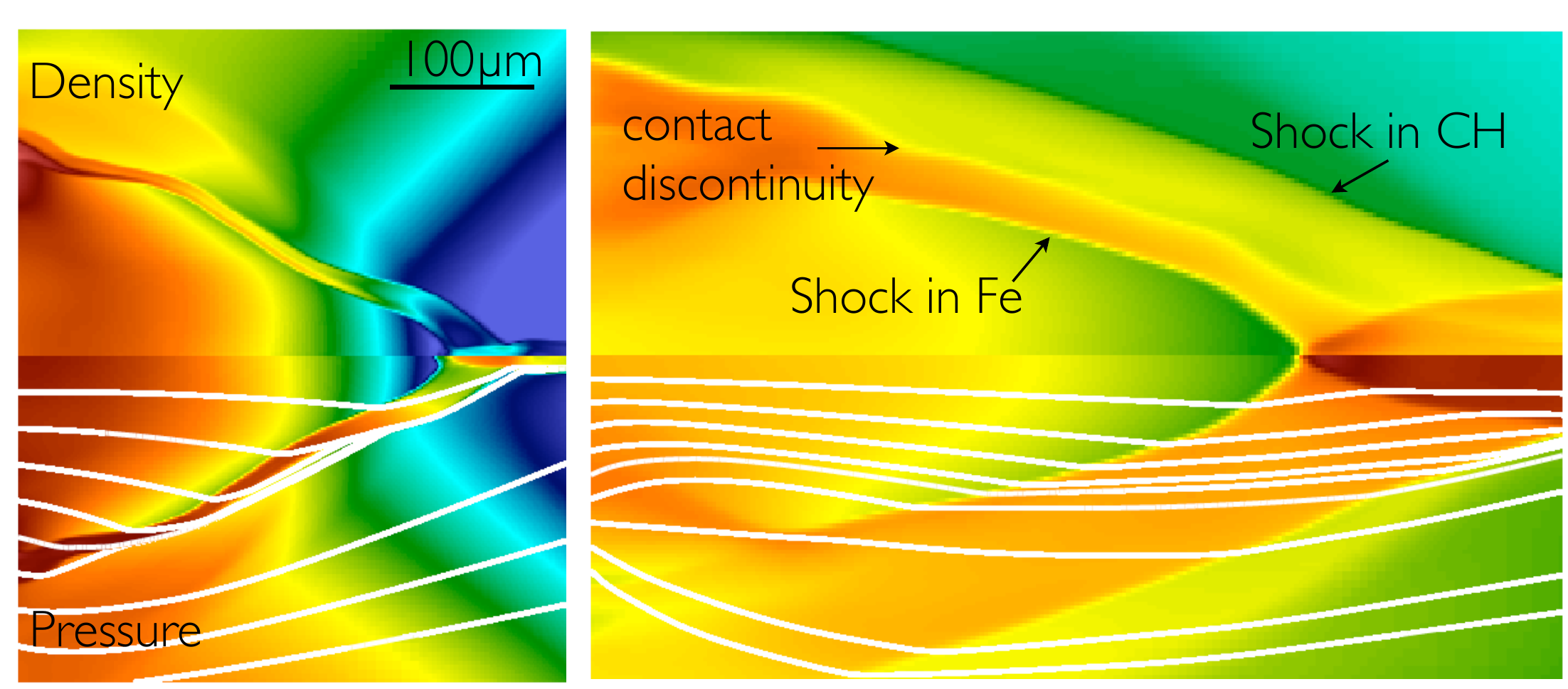} 
\caption{Streamlines on density and pressure maps for simulations at 8 ns (a) and 20 ns (b) showing the ``focusing"  of the iron on the axis as it strikes the shock. The transmitted shock in Fe, the reflected shock in CH and the contact discontinuity in-between are clearly distinguishable in the density map at 20 ns.  Color scheme is same as in Fig.  \ref{Simulations}}
\label{fig: streamlines} 
\end{center} 
\end{figure}

{\it Numerical Simulations and Physical Interpretation-} Using the FLASH 
 multi-physics AMR (Adaptive Mesh Refinement) code \cite{Fryxell2000}  (recently  extended to include high energy density physics capabilities \cite{Orban2013}), 
we  have simulated the interaction of the iron in with the surrounding plastic in 2-D.
We used the un-split 3-temperature HD solver with the energy deposition module and a radiation transfer modeled by multi-group diffusion with 32 groups. Iron and plastic shock breakout times  measured by the rear-side self-emission diagnostics, 
are used to calibrate the laser intensity in the simulations. 
The reliability of the simulations up to 55 ns is also verified by interferometry and shadowgraphy data.  
At later times, the cumulative incertitude associated with  equations of state, opacities, conduction models, species mixing, etc, limits the accuracy of the numerical results.
 Detailed  modeling of the plasma parameters is beyond the present scope, but the present simulations do very much help to convey the global flow dynamics and physical processes  \\
\indent Fig. \ref{Simulations}a shows   density and  pressure maps at different times,  with the corresponding synthetic radiographs (Fig. \ref{Simulations}b-f). 
These panels  can be  directly compared to the experimental results.
Generally, there is substantial agreement with the data shown in Fig. \ref{radiography:time_series}b-h
with respect to the presence and time evolution of the iron jet.  
All of the different evolutionary phases are seen in the simulations: confinement;   focusing  with a convergence point (Fig. \ref{Simulations}c-d); and  long lasting  collimation  (Fig. \ref{Simulations}e-f). Simulations also show that the iron collimation initiates at a shock wave from the collision between supersonic CH and Fe plasmas.  Signatures of the shock are seen in the simulated density and pressure maps, which indicate three density discontinuities and  two pressure jumps, corresponding to the inner and outer density features (Fig. \ref{fig: streamlines}). These features correspond to  a transmitted shock in Fe, a reflected shock in CH and a contact discontinuity between them. 
 The shock is also seen in the simulated X-ray radiographs as a stronger absorption layer and corresponds to the high density shocked iron, already observed  in the experimental data.
Together with  the increase in the emitted radiation recorded at the iron/plastic boundary  from the rear side self-emission, 
  the experimental measurements are  consistent with the  picture revealed by simulations.
  

The presence of the shock and its shape reveals the  dynamics and collimation of the iron flow. The shock shape is determined by the relative expansion of the CH and Fe plasmas. In our case, a converging conical shock is generated in Fe, since the CH plasma forms earlier and expands farther than the Fe. As the Fe expands and strikes the shock surface obliquely, only the normal component of the post-shock velocity is reduced, so the  shock  marks  the locus at which the Fe flow vector  focuses toward the axis. 
 (Fig. \ref{fig: streamlines}). 
This mechanism was previously  identified analytically in astrophysical studies \cite{Sanders} and later in HD simulations \cite{icke93,Frank1994,Frank1996b}. It  is  called ``shock-focused inertial confinement" (SFIC)  and may help  collimate  flows from  Young Stellar Objects (YSO) and PN,  particularly when cooling is added \cite{Mellema1997}. Our results give the first experimental confirmation of this scenario, without cooling.

{\it  Astrophysical Relevance-}
The importance of the  experiment  is bolstered  when 
   experimental  parameters  correspond to  those of specific astrophysical systems \cite{Ryutov99}.
The characteristic experimental parameters for the iron flow are shown in the first column  of table \ref{tab:table1}.  They indicate a highly collimated (AR$\sim5$), supersonic flow (M$\sim$10) in a pure HD regime where radiative ($\chi\gg1$) and microphysical conductive ($Pe\gg1$) effects are negligible. 
Table \ref{tab:table1} also shows  representative parameter regimes for  YSOs \cite{Hartigan1993,reipurth01,arce02}, 
AGN \cite{Ferrari1998,Tombesi2012}, and PPN  \cite{bujarrabal,Balick,blackmanlucchini,witt09,rizzo2013},   near  the inner collimation scales  of these jets.   
 Note that $Pe>>1$ in all cases  so that microphysics of thermal conduction does no affect the bulk dynamics.
     YSOs are the most similar  to the experiments,  except for their  cooling. 
     \\
     \indent   
   For PPN,  the young jets of low density seem to interact with the denser wind of the post-AGB star \cite{witt09}, resulting in a density ratio $<1$. AGN jets are also of lower density than their surrounding wide angle winds and  they are relativistic, differing in those respects from the experiments.
Nevertheless,  jet collimation in the experiment arises because the momentum of the outer outflow can redirect that of the inner outflow, and this requirement  would be the same  regardless of the density ratio of whether the flows are relativistic.   So the  nested wind structure and basic principles of inertial collimation still apply  to PPN and AGN, but the specific predictions for shock location and geometry could be different.

\textit{Conclusions-} Motivated by astrophysical contexts where jets are associated with accretion engines, we have established an experimental platform to study the collimating interaction between high Mach outflows. We have experimentally confirmed the efficacy of inertial mechanisms in producing highly collimated outflows in an HD regime, similar to jets from YSOs, but also relevant for systems such as AGN and PPN.
Most importantly, we have experimentally verified the SFIC mechanism suggested in previous  astrophysical studies, exemplifying the contribution  that such laboratory experiments can bring to the enterprise of astrophysics.  
Further insights on the jet collimation paradigm can be accessed by combining inertial with magnetic and radiative effects in future experiments.
\medskip

\medskip
{\it Acknowledgments-}
We gratefully acknowledge  support of the technical staff at the LULI 2000 laser facility and H. Nakatsutsumi for target fabrication. 
The FLASH code  was developed in part by the U.S. DOE and NSF-funded Flash Center for Computational Science at the Univ. of Chicago. This work was performed using HPC resources from GENCI-IDRIS (Grant 2013-i2013057066). EB acknowledges NSF grant AST-1109285.


\newpage 

\noindent 
FIG. 1: Schematic  of the experimental setup, focal spot (measured) and target geometry. The laser energy distribution is modeled by means of a hybrid phase plate (HPP) resulting in a central disc and an outer ring.  The nested target fits this geometry, consisting in a central iron (Fe 15$\mu$m) disc and and outer plastic (CH 15$\mu$m) ring on a common CHAl pusher. The laser hits the  front side of the target and the plasma emanates from the rear..\\ 

\noindent 
FIG. 2: Typical time resolved rear side self-emission data for a) the iron disc only and b) the complete nested target. \\ 

\noindent 
FIG. 3: X-ray radiography of the iron flows from a) the central disk alone, where the outer ring was blocked on the laser side and b)-h) from the complete nested target. 
In this case, the temporal evolution shows the formation of a collimated flow being stable up to very long delays (80 ns) and reaching mm sizes. In the insertion at 100 ns  we have enhanced the contrast.  \\ 

\noindent 
FIG. 4: Upper panel: simulated density (upper) and pressure (lower) maps at different time delays. Lower panel: corresponding synthetic radiographs obtained using the simulated density map and the experimental X-ray spectrum.  \\ 

\noindent 
FIG. 5: Streamlines on density and pressure maps for simulations at 8 ns (a) and 20 ns (b) showing the ``focusing"  of the iron on the axis as it strikes the shock. The transmitted shock in Fe, the reflected shock in CH and the contact discontinuity in-between are clearly distinguishable in the density map at 20 ns.  Color scheme is same as in Fig. 4 \\


\noindent
TABLE 1: Parameter values for the experiment and typical astrophysical cases on scales where collimation occurs. The index j stands for ``jet"  while the index a for ``ambient" , represented in the experiment by the Fe and CH respectively . V is the velocity, $c_s$ the sound velocity, l the longitudinal length, r the radial extension, $\rho$ the density, $t_{rad}$ is the cooling time, while $t_{hydro}$ is the hydrodynamic one. $\chi$ is the thermal diffusivity, calculated as in \cite{Ryutov99}. c is the speed of light
 

\newpage

\newpage 
\begin{table}
\caption{Experimental vs.  astrophysical parameters  on scales where collimation occurs. The indices j and a
refer to "jet"  and ``ambient" , represented in the experiment by the Fe and CH respectively . V is  velocity, $c_s$ is sound speed, l is the longitudinal length, $r$ the radial extension, $\rho$ the density, $t_{rad}$ is the cooling time, $t_{hydro}$ is the hydrodynamic time. $\chi$ is the thermal diffusivity, as in \cite{Ryutov99}. $c$ is the speed of light}
\label{tab:table1}
\begin{ruledtabular}
\begin{tabular}{lllll}
Parameter  &Lab. &  YSO  & PPN & AGN \\
\hline          
 collimation  scale  & $ 1$mm  & $10^{-3} pc$ &$<0.01$pc & $$0.1pc\\
Int. Mach $M_{int}$=$V_j/c_{s,j}$ & 5-10  & $>10$ &$>10 $&$>10$\\
		 Ext. Mach $M_{ext}$=$V_j/c_{s,a}$ & 5-10  & $>10$ &$>10$ & $> 10$\\
                    aspect ratio $AR=l_j/r_j$ & 5 & ~10&~10&$>10 $\\
                  density ratio $\eta=\rho_j/\rho_a$ & 5-10&10 &$< 1$& $<<1$ \\
            	 Cooling $\chi=t_{rad}/t_{hydro}$  & $100$  & $<1$ &$<1$&$>>1$\\     
                    Peclet $Pe=\rho r V_j/\chi$  & $10^4$  & $>>1$ &$>>1$&$>>1$ \\   
                    $\beta=V_j/c$& $10^{-4}$& 10$^{-3}$& $ 10^{-3}$& 0.9-0.99\\
\end{tabular}
\end{ruledtabular}
\end{table}

\end{document}